\documentclass{ws-mplb}

\begin{document}

\markboth{Z. Papp and S. Moszkowski}{ Two- and three-alpha systems with nonlocal potential}

%
\catchline{}{}{}{}{}
%

\title{   Two- and three-alpha systems with nonlocal potential }

\author{Z. Papp}

\address{Department of Physics and Astronomy, California State University Long Beach, \\
Long Beach, 
California 
\\
zpapp@csulb.edu}

\author{S. Moszkowski}

\address{Department of Physics and Astronomy, University of California Los Angeles, \\
 Los Angeles, 
California }

\maketitle

\begin{history}
\received{(\today)}
\revised{(\today)}
\end{history}

\begin{abstract}
Two body data alone cannot determine the potential uniquely, one needs three-body data as well.
A method is presented here which simultaneously fits local or nonlocal potentials to two-body and three-body observables. The interaction of composite particles, due to the Pauli effect and the indistinguishability of
the constituent particles, is genuinely nonlocal. As an example, we use a Pauli-correct nonlocal fish-bone type optical model for the $\alpha-\alpha$ potential and derive the fitting parameters such that it reproduces the two-$\alpha$ and three-$\alpha$ experimental data. 
\end{abstract}


\section{Introduction}

The aim of inverse scattering methods is to determine the operator from the spectrum.
If we assume nonrelativistic dynamics, the inverse problem boils down to determining the potential.
Mathematics alone, however, cannot determine the potential uniquely. A general potential operator, taken in coordinate representation, is a nonlocal potential
\begin{equation}
\langle r | V | r' \rangle = V(r,r')~.
\end{equation}
So, if we write the Schr\"odinger equation in the form
\begin{equation}
\left[-\frac{\hbar^2}{2m}\left( \frac{d^2}{dr^2}-\frac{l(l+1)}{r^2}\right)+ V_l(r)\right]\psi(r)=E\psi(r),
\end{equation}
where $m$ is the mass of the particle and $l$ is the angular momentum, we assume implicitly that the potential is local
\begin{equation}
V_l(r,r')=V_l(r) \delta(r-r').
\end{equation}
This is a serious approximation since, for example the interaction between composite particles, due to the internal motion of the constituents, is a genuine nonlocal potential. 

The physical input to the inverse scattering procedure is the bound-state energy and the phase shift, or something equivalent. The bound state wave function is square integrable and 
\begin{equation}
\psi(r)\sim \exp(-\kappa r), 
\end{equation}
as $r\to\infty$, where $\kappa=\sqrt{2m/\hbar^2 |\epsilon|}$ and $\epsilon$ is the bound-state energy. If the potential is of  short-range type the scattering wave function at large distances looks like
\begin{equation}
\psi(r) \sim \exp(-i k r + l {\pi}/{2} + \delta),
\end{equation}
where $k=\sqrt{2m/\hbar^2 E}$, $E$ is the energy and $\delta$ is the phase shift. 
So, the spectrum of the operator is contained in the asymptotic behavior of the wave function.
Or, other way round, the spectrum determines only the asymptotic behavior of the wave function.
The internal part is arbitrary, at least to some extent.

To determine the potential unambiguously we need to know the internal part of
the wave function as well. In order that we gain some information about the internal part of the wave function we may add a third particle to the system. The third particle feels not only the asymptotic part, but also the whole wave function. So, three-body observables are sensitive not only to the asymptotic part, but also to the whole two-body wave function. Therefore, to determine the potential uniquely, we need to study both two-body and three-body systems. 

The aim of this work is to present a quantum mechanical approximation method which can handle
realistic local or nonlocal potentials with long range Coulomb tail in two- and three-body dynamics. 
For this purpose we developed an integral equation approach. For two-body systems we solve the Lippmann-Schwinger integral equation, and for three-body systems we solve the Faddeev integral equations. Here we sketch the solution method, the details are given in preceding publications\cite{papp1,papp2}.

An argument against using nonlocal potentials is that it is too general and by using them we open 
a "Pandora's box". However, we often have some information about the internal structure of the composite particles which provides some framework for the nature of the nonlocality. Composite particles often made of identical elementary fermions. If two composite particles gets closer, the Pauli principle prevent the identical constituents from occupying the same quantum state. 
This is the main source of the nonlocality of the interaction at short distances. There are several cluster model inspired effective potentials which has nonlocal parts.
Probably the most general and best founded model is the fish-bone model \cite{schmid}. This approach takes into account not only the complete Pauli blocking, but also it allows that some states are partly Pauli forbidden or Pauli suppressed. 

In this work we examine the $\alpha-\alpha$ interaction in the fish-bone model. We determine the $\alpha-\alpha$ potential such that it reproduces the $S$, $D$ and $G$ wave phase shifts up
to $20$ MeV, the famous the $E=91$ KeV, $l=0$ $\alpha-\alpha$ resonant state, and the 
three-$\alpha$ binding energy. 

In Sec.\ II we present the Coulomb-Sturmian separable expansion approach to the two-body
Lippmann-Schwinger equation\cite{papp1}. In Sec.\ III we extend the method for solving the Faddeev integral equations of the three-body problem\cite{papp2}. The Faddeev equations with Coulomb potentials are rather complicated, so we restrict ourself to bound states and identical particles. Then, in Sec.\ IV, we reexamine the two-$\alpha$ and the three-$\alpha$ system and determine a new fish-bone-type $\alpha-\alpha$ potential which provides a good description to the two-$\alpha$ and three-$\alpha$ observables.

\section{Two-body problem}
\subsection{Coulomb-Sturmian separable expansion approach to Lippmann-Schwinger equation}

In this section we recapitulate the solution of the Lippmann--Schwinger 
equation by using the Coulomb-Sturmian separable expansion method. 
We suppose that the total Hamiltonian $h_l$ can be split into two terms  
\begin{equation}
h_l=h_l^{\rm C} + v_l \ ,
\end{equation}
where $v_l$ is an asymptotically irrelevant short-range potential and $h_l^{\rm C}$ 
is the Coulomb Hamiltonian
\begin{equation}
 h^{\rm C}_l=-\frac{\hbar^2}{2m}\left(\frac{\mbox{d}^2 }{\mbox{d} r^2}
- \frac{l(l+1)}{ r^2}\right) + \frac{Z}{ r}\ ,
\label{coulham}
\end{equation}
where $m$ is the reduced mass, $l$ is the angular momentum and $Z$ is the strength of the Coulomb potential.

We denote the Coulomb Green's operator by $g_l^{\rm C}(z)=(z-h^{\rm C}_l)^{-1} $
and the Green's operator of the total Hamiltonian by
$g_l(z)=(z-h_l)^{-1}$.
They are related via the resolvent relation 
\begin{equation}
g_l(z)=g_l^{\rm C}(z)+g_l^{\rm C} (z)v_l g_l(z)\ .
\label{resolvent}
\end{equation}
The scattering wave function $|\psi^{(+)}_l \rangle$ 
satisfies the inhomogeneous Lippmann--Schwinger 
equation 
\begin{equation}
|\psi^{(+)}_l  \rangle =|\varphi^{\rm C}_l \rangle 
+g_l^{\rm C}(E + {\rm i}0) v_l |\psi^{(+)}_l \rangle \ ,  
\label{LS}
\end{equation}
while the bound- and resonant-state  wave function
satisfies the homogeneous Lippmann--Schwinger  equation
\begin{equation}
|\psi_{l }  \rangle = g_l^{\rm C}(E ) v_l |\psi_{l }\rangle  
\label{LS1}
\end{equation}
at negative real and complex $E$ energies, respectively.

We solve these equations in a unified way by approximating
only the short range potential $v_l$. We use  Coulomb-Sturmian (CS) functions
\begin{equation}
\langle r\vert n l ;b  \rangle 
= \left( \frac{n!}{(n+2l+1)!} \right)^{1/2} \ 
\exp(-b r) (2b r)^{l+1} L_n^{(2l+1)}(2b r)\ , 
\label{csf}
\end{equation}
where $L$ denotes the Laguerre polynomial.
The CS functions form a bi-orthonormal basis. With 
$\langle r\vert \widetilde{n l;b} \rangle \equiv \langle r\vert n l;b \rangle/r$  we have
\begin{equation} \label{orto}
\langle \widetilde{ n'l;b }  \vert nl;b \rangle 
= \langle n'l;b  \vert \widetilde{ n l;b }  \rangle = \delta_{n n'}.
\end{equation}
To approximate the potential operator we use a ''skew'' form of the completeness relation
${\bf 1}_{l}=\lim_{N\to\infty} {\bf 1}^{N}_{l}$, where 
\begin{equation}
{\bf 1}^{N}_{l}= \sum _{n,m=0}^{N }|\widetilde{n l;b_g}\rangle 
(\underline{O}^{-1})_{n m}
\langle m l; b_v| 
\label{1N}
\end{equation}
with $(\underline{O})_{m n}=
(\langle m l;b_v |\widetilde{n l;b_g}\rangle)$. It is obvious that this double sum 
is also a possible expression for unity. If $b_{v}=b_{g}$, Eq.\  (\ref{1N}) falls back to the usual completeness relation. By adopting different values for $b_{v}$ and $b_{g}$ one can achieve a faster convergence in the separable approximation\cite{adhikari}. 

In Eqs.\ (\ref{LS}) and (\ref{LS1}) the term $v_l |\psi_{l }\rangle$ is square integrable, and belongs to the Hilbert space.    Therefore, it can be approximated by
\begin{equation}
v_l |\psi_{l }\rangle   \approx  {\bf 1}_{l}^{N} v_l |\psi_{l }\rangle   
\approx {\bf 1}^{N}_{l} v_l  {\bf 1}^{N}_{l} |\psi_{l }\rangle = v_l^N |\psi_{l }\rangle= 
\sum_{n,n'}^N
|\widetilde{n l;b_g}\rangle  \;
\underline{v}_{n n'} \;\mbox{} \langle \widetilde{n' l;b_g}|\psi_{l }\rangle \ 
\label{sepfe2b}
\end{equation}
with finite $N$, where 
\begin{equation}
 \underline{v}_{n n'} =\sum_{m,m'=0}^N 
 (\underline{O}^{-1})_{n m} \langle m l;b_v | v_l| m' l;b_v \rangle 
(\underline{O}^{-1})_{m' n'}~. 
\label{v2b}
\end{equation}
In general, the matrix elements $\langle m l;b_v | v_l| m' l;b_v \rangle$ 
have to be calculated numerically. Note that the potential is sandwiched
between CS states with parameter $b_v$ and the potential operator becomes 
a linear combination of CS ket-bra operators 
$| nl;b_g \rangle \langle n'l;b_g |$.
This approximation is called separable expansion because the potential operator after approximation
appears in the form 
\begin{equation}
\langle r | v^N_l| r' \rangle = \sum_{n, n' =0}^N
\langle r |\widetilde{nl;b_g}\rangle  \;
 \underline{v}_{n n'} \;\mbox{} \langle \widetilde{n'l;b_g}| r' \rangle\ , 
\end{equation}
i.e.\  the dependence on $r$ and $r'$ is separated. 

With this  separable potential Eqs.\ (\ref{LS}) and  (\ref{LS1}) become
\begin{equation}
|\psi^{(+)}_l \rangle =|\varphi^{\rm C}_l \rangle + \sum_{n,n^\prime =0}^N
g^{\rm C}_l (E + {\rm i}0)  |\widetilde{n l;b_g}\rangle  \;
\underline{v}_{n n'} \;\mbox{} \langle \widetilde{n'l;b_g}
|\psi^{(+)}_l \rangle\ ,  
\label{LSapp1}
\end{equation}
and 
\begin{equation}
|\psi_{l } \rangle = \sum_{n,n^\prime =0}^N
g_l^{\rm C}(E ) |\widetilde{n l;b_g}\rangle  \;
 \underline{v}_{n n'}\;\mbox{} \langle \widetilde{n' l;b_g}
|\psi_{l} \rangle\ ,  
\label{LSapp2}
\end{equation}
respectively. To derive equations for the unknown
coefficients $\underline{\psi}_l^{(+)} =
\langle \widetilde{n'l;b_g}|\psi_l^{(+)} \rangle$ and 
$\underline{\psi }_l =\langle \widetilde{n'l;b_g}|\psi_{l} \rangle$, 
we have to act with states $\langle \widetilde{n'' l;b_g}|$
from the left. Then, the following inhomogeneous and homogeneous 
algebraic equations are obtained for scattering and bound-state 
problems, respectively: 
\begin{equation}
[ (\underline{g}^{\rm C}_l (E + {\rm i}0))^{-1}-
\underline{v}_l ]\underline{\psi}_l^{(+)}=
(\underline{g}^{\rm C}_l (E + {\rm i}0))^{-1}
\underline{\varphi}_l^{\rm C},  \label{eq18a}
\end{equation}
and
\begin{equation}
\lbrack (\underline{g}_l^{\rm C} (E))^{-1}-
\underline{v}_l]\underline{\psi }_l= 0\ .
  \label{eq18b}
\end{equation}
Here we have $\underline{g}_{l}^{\rm C}=\langle \widetilde{nl;b_g}|g_{l}^{\rm C}|\langle \widetilde{n'l;b_g}\rangle$, the matrix elements of the Coulomb Green's operator, and
the overlap of the CS and Coulomb functions $\underline{\phi}_l^{\rm C} =\langle \widetilde{n'l;b_g}|\phi_{l}^{\rm C} \rangle$. The reason that we have adopted the CS basis is that, in that basis,  these quantities can be calculated analytically\cite{papp1,fhp}.
 The homogeneous equation (\ref{eq18b}) is solvable if and only if 
\begin{equation}
\label{det}
\det [(\underline{g}_l^{\rm C}(E))^{-1}-\underline{v}_l]=0 
\end{equation}
holds, which is an implicit nonlinear equation for the bound- and
resonant-state energies.  
As far as the scattering states are concerned, the solution of (\ref{eq18a}) 
provides the overlap $\langle \widetilde{nl;b_g}|\psi_l \rangle$. 
From this quantity
any scattering information can be inferred, for example the Coulomb-modified 
scattering amplitude reads
\begin{equation}
\label{scamp}
 a_l = \langle 
\varphi^{\rm C (-)}_l | v_l | \psi^{(+)}_l  \rangle = 
\underline{\varphi}_l^{\rm C (-)}
\underline{v}_l
 \;\underline{\psi}_l^{(+)} \ ,
\end{equation}
which is related to the Coulomb-modified short-range phase shift $\delta_l$ through
\begin{equation}
a_l= \frac{1}{k}\exp( \mbox{i} (2 \eta_l +  \delta_l)) \sin \delta_l \ ,
\end{equation}
where $\eta_l=\arg \Gamma(l+\mbox{i}\gamma+1)$ is the Coulomb phase shift.

In this approach, the potential enters into the procedure through its CS matrix elements. 
So, the method works for any potential, local or nonlocal, energy dependent, complex, etc., as long as we can evaluate its CS matrix elements somehow. Numerical
examples show that the method is also very efficient\cite{papp1}. Having the optimal values for 
$b_{v}$ and $b_{g}$, one need about $N=15-20$ basis states to achieve $6-7$ digits accuracy for the bound state energies and for the phase shifts over the entire spectrum.

\section{Three-body problem}

\subsection{The Faddeev equations}

Here we consider three identical particles interacting with repulsive Coulomb-like potentials and we restrict ourself to bound states. The method is applicable also for attractive Coulomb potentials and for scattering and resonant states as well\cite{papp2}. The  Hamiltonian is given by
\begin{equation}
H=H^0 + v_1 + v_2 + v_3,
\label{H}
\end{equation}
where $H^0$ is the three-body kinetic energy 
operator and $v_i$ denotes the long-range Coulomb-like potential of each subsystem $i=1,2,3$. 
We use the usual configuration-space Jacobi coordinates:  $x_1$ is the distance
between the pair $(2,3)$ and $y_1$ is the
distance between the center of mass of the pair $(2,3)$ and the particle $1$.

A Coulomb potential modifies the character of the asymptotic motion, therefore it should be treated very much like the kinetic energy operator. We split the potential into two parts, a short-range and a Coulomb part
\begin{equation}
v_i =v^{(s)}_i +v^{\rm C}_i.
\label{pot}
\end{equation}
We define 
the long-range Hamiltonian by
\begin{equation}
H^{(l)} = H^0 + v_1^{\rm C}+ v_2^{\rm C}+ v_3^{\rm C},
\label{hl}
\end{equation}
and the three-body Hamiltonian takes the form
\begin{equation}
H = H^{(l)} + v_1^{(s)}+ v_2^{(s)}+ v_3^{(s)}.
\label{hll}
\end{equation}
This Hamiltonian looks like an ordinary three-body Hamiltonian with 
short-range interactions. We solve the 
Schr\"odinger equation
\begin{equation}
H|\Psi\rangle =E|\Psi\rangle
\end{equation}
by using the Faddeev method. We split the wave function
into three components
\begin{equation}\label{psi12}
|\Psi \rangle =| \psi_1 \rangle + | \psi_2 \rangle + | \psi_3 \rangle,
\end{equation}
and the components should satisfy the 
set of differential equations 
\begin{subequations}\label{fm3compd}
\begin{eqnarray}
(E-H_{1}^{(l)}) | \psi_1 \rangle &= &  v_1^{(s)} (| \psi_2  \rangle + | \psi_3  \rangle ) \\
(E-H_{2}^{(l)}) | \psi_2 \rangle &= &  v_2^{(s)} (| \psi_1 \rangle + | \psi_3  \rangle ) \\
(E-H_{3}^{(l)}) | \psi_3 \rangle &= &  v_3^{(s)} (| \psi_1 \rangle + | \psi_2  \rangle ),
\end{eqnarray}
\end{subequations}
where $H_{i}^{(l)}=H^{(l)}+v_{i}^{(s)}$.
By adding these three equations we get back the original Schr\"odinger equation.

By inverting the left hand side of (\ref{fm3compd}) we get a set of integral eqautions\begin{subequations}\label{fm2comp}
\begin{eqnarray}
| \psi_1 \rangle &=  & G_1^{(l)}(E) v_1^{(s)} (| \psi_2  \rangle  + | \psi_3  \rangle) \\
| \psi_2 \rangle &=  & G_2^{(l)}(E) v_2^{(s)} (| \psi_1  \rangle  + | \psi_3  \rangle) \\
| \psi_3 \rangle &=  & G_3^{(l)}(E) v_3^{(s)} (| \psi_1  \rangle  + | \psi_2  \rangle),
\label{fm2c2}
\end{eqnarray}
\end{subequations}
where $G_i^{(l)}(E)=(E-H_{i}^{(l)})^{-1}$.

If particles $1$, $2$ and $3$ are identical,
then $\psi_{1}$, in its natural Jacobi coordinate system $\{x_{1},y_{1}\}$, looks like $\psi_{2}$ in its natural Jacobi coordinate system $\{x_{2},y_{2}\}$ and $\psi_{3}$ in its natural Jacobi coordinate system $\{x_{3},y_{3}\}$.
On the other hand, by interchanging particles $2$ and $3$ we have
\begin{equation}
{\mathcal P}_{23} | \psi_1 \rangle = p | \psi_1 \rangle,
\end{equation}
where  $p=1$ for bosons and $p=-1$ for fermions.
Building this information into the formalism we arrive at
a single integral equation
\begin{equation} \label{fmp}
| \psi_{1} \rangle =  2 G_1^{(l)} v_1^{(s)} {\mathcal P}_{123}
| \psi_{1} \rangle,
\end{equation}
where ${\mathcal P}_{123}={\mathcal P}_{12}{\mathcal P}_{23}$ is the operator for cyclic permutation of all three particles 
${\mathcal P}_{123} |\psi_{1}\rangle =| \psi_{2}\rangle$.
We should notice that so far no approximation has been made, and even though 
this integral equation has only one component, yet it
gives a full account for the asymptotic and symmetry properties of the system.

\subsection{Coulomb-Sturmian expansion}

Since the three-body Hilbert space is a direct product of two-body
Hilbert spaces, an appropriate basis is the bipolar basis, which
can be defined as an
angular-momentum-coupled direct product of the two-body bases, 
\begin{equation}
| n \nu  l \lambda ; b_x b_y \rangle_\alpha =
 | n  l ; b_x \rangle_\alpha \otimes |
\nu \lambda ;b_y\rangle_\alpha, \ \ \ \ (n,\nu=0,1,2,\ldots),
\label{cs3}
\end{equation}
where $| n  l ;b_x\rangle_\alpha$ and $|\nu \lambda ;b_y\rangle_\alpha$ 
are associated
with the coordinates $x_\alpha$ and $y_\alpha$, respectively.
With this basis the completeness relation
takes the form (with angular momentum summation implicitly included)
\begin{equation}
{\bf 1} =\lim\limits_{N\to\infty} \sum_{n,\nu=0}^N |
 \widetilde{n \nu l \lambda ;b_x b_y} \rangle_\alpha \;\mbox{}_\alpha\langle
{n \nu l \lambda ;b_x b_y} | =
\lim\limits_{N\to\infty} {\bf 1}^{N}_\alpha,
\end{equation}
where $\langle x y | \widetilde{ n \nu l \lambda;b_x b_y}\rangle = 
\langle x y | { n \nu l \lambda; b_x b_y}\rangle/(x y)$.

Similarly to the two-body case, $v_1^{(s)} {\mathcal P}_{123} | \psi_{1} \rangle$ is square integrable, therefore we can approximate 
\begin{eqnarray}
v_1^{(s)}{\mathcal P}_{123} | \psi_{1} \rangle  & = & \lim_{N\to\infty} 
{\bf 1}^{N}_1  v_1^{(s)}{\mathcal P}_{123}  {\bf 1}^{N}_1 | \psi_{1} \rangle 
\approx  {\bf 1}^{N}_1 v_1^{(s)} {\mathcal P}_{123} {\bf 1}^{N}_1  | \psi_{1} \rangle
 \nonumber \\
& \approx & 
 \sum_{n,\nu ,n', \nu'=0}^N
|\widetilde{n\nu l \lambda,b_x b_y}\rangle_1 \; \underline{V}_1
\;\mbox{}_1 \langle \widetilde{n' \nu' l' \lambda';b_x b_y}   | \psi_{1} \rangle ,  \label{sepfe}
\end{eqnarray}
where $\underline{V}_{1}  = \underline{v}_{1}^{(s)} \langle \widetilde{nl;b_{x} } | \widetilde{n'l;b_{x}}\rangle \;
\mbox{}_1 \langle n\nu l \lambda; b_x b_y  |n' \nu' l' \lambda' ;b_x b_y \rangle_2$.
The completeness of the CS basis guarantees the convergence of the expansion
with increasing $N$ and angular momentum channels.

Now, by applying the bra $\langle \widetilde{ n'' \nu'' l'' \lambda'';b_x b_y}|$
on Eq.\ (\ref{fmp}) from the left, the solution of the homogeneous
Faddeev equation
turns into the solution of a matrix equation for the component vector
$\underline{\psi}_{1}=
 \mbox{}_1 \langle \widetilde{ n\nu l\lambda; b_x b_y} | \psi_{1}  \rangle$
\begin{equation}
 \underline{\psi}_{1} =  2 \underline{G}_1^{(l)}  
\underline{V}_1   \underline{\psi}_{1} , \label{fn-eq1sm}  
\end{equation}
where 
\begin{equation}
\underline{G}_1^{(l)}=\mbox{}_1 \langle \widetilde{
n\nu l\lambda;b_x b_y} |G_1^{(l)}|\widetilde{n' \nu' l' \lambda';b_x b_y}
\rangle_1.
\end{equation}
This homogeneous algebraic equation is solvable if only if
the determinant is zero:
\begin{equation}
\det [  (\underline{G}_1^{(l)})^{-1}  -2 \underline{V}_1] = 0. \label{homdet}  
\end{equation}

The operator $\underline{G}_1^{(l)}$ is the resolvent of a complicated three-body Coulomb
Hamiltonian. This Hamiltonian, however can support only one kind of asymptotic
channel, when particles $2$ and $3$ are close and particle $1$ is at infinity.
The asymptotic Hamiltonian of $H_{1}^{(l)}$ is 
\begin{equation}\label{htilde}
\widetilde{H}_{1}=H^{0}+v_{1}+u_{1},
\end{equation}
where $u_{1}^{\rm C}$ is the channel Coulomb potential. This potential is the asymptotic part
of $v_{2}^{\rm C}+v_{3}^{\rm C}$ as particle $1$ is separated from the pair $(2,3)$. 
It looks like the charges of particles $2$ and $3$ are concentrated in their center of mass
\begin{equation}
u_{1}^{\rm C}=Z_{1}(Z_{2}+Z_{3})/y_{1},
\end{equation}
where $Z_{i}$ is the charge of the particles.

For Hamiltonians which have  one kind of asymptotic Hamiltonian, 
a single Lippmann-Schwinger
equation provides a unique solution. Thus
\begin{equation}
G_1^{(l)}(z)=\widetilde{G}_1(z) + \widetilde{G}_1 (z) U^{(l)}_1 G_1^{(l)}(z),
\label{LSass}
\end{equation}
where  $\widetilde{G}_1(z)=(z-\widetilde{H}_{1})^{-1}$ and $U_{1}^{(l)}=v_{2}^{\rm C}+v_{3}^{\rm C}-u_{1}^{\rm C}$. 
To solve the Faddeev equation, we need $G_1^{(l)}$ between finite number of square integrable
CS basis states. We make a separable approximation on $U_{1}$, and with the help of matrix elements 
$\underline{U}_{1} =
 \mbox{}_1\langle n\nu l \lambda;b_x b_y | U_1 | 
 n' \nu' l' \lambda';b_x b_y \rangle_1$,
we get
\begin{equation}
(\underline{G}^{(l)}_1)^{-1}= 
(\underline{\widetilde{G}}_1)^{-1} - \underline{U}_1,
\label{gleq}
\end{equation}
where $
\underline{\widetilde{G}}_{1} =
 \mbox{}_1\langle \widetilde{n \nu l \lambda;b_x b_y} | 
 \widetilde{G}_1 | \widetilde{ n' \nu' l' \lambda';b_x b_y} \rangle_1.  
$
 This later matrix element of $U_{1}$ can always be evaluated numerically.

To calculate the matrix elements $\underline{\widetilde{G}}_{1}$ we utilize the Dunford-Taylor functional calculus.
If $h$ is a selfadjoint operator, then an analytic function $f$ of $h$ is given by
a contour integral 
\begin{equation}
f(h)=\frac{1}{2\pi i}\oint_{C} dz\: f(z) (z-h)^{-1},
\end{equation}
where the contour $C$ goes around the spectrum of $h$ such that $f$ is analytic on the area encircled by $C$.

The three-particle free Hamiltonian
can be written  as a sum of two-particle free Hamiltonians 
\begin{equation}
H^0=h_{x_1}^0+h_{y_1}^0.
\end{equation}
Thus the Hamiltonian $\widetilde{H}_1$ of Eq.\ (\ref{htilde}) 
appears as a sum of two two-body Hamiltonians acting on different coordinates 
\begin{equation}
\widetilde{H}_1 =h_{x_1}+h_{y_1},
\end{equation}
with $h_{x_1}=h_{x_1}^0+v_1^C(x _1)$ and $h_{y_1}=h_{y_1}^0+u_{1}^{\rm C}(y_{1)}$, 
which, of course, commute.
The Green's operator 
\begin{equation}
\widetilde{G}_{1}(E)=(E-h_{x_{1}}-h_{y_{1}})^{-1}
\end{equation}
is a function of the selfadjoint operator $h_{y_{1}}$. So,
\begin{equation}\label{contourint}
\widetilde{G}_{1}(E)=\frac{1}{2\pi i}\oint_{C} dz\: (E-h_{x_{1}}-z)^{-1} (z-h_{h_{1}})^{-1}=
\frac{1}{2\pi i}\oint_{C} dz\: g_{x_{1}}(E-z)\: g_{y_{1}}(z),
\end{equation}
i.e.\ we can calculate the three-body resolvent as a convolution integral of two-body resolvent operators.

In this work we concentrate on bound states of the three-$\alpha$ system. Therefore  $E<0$, $h_{y_{1}}$ is a repulsive Coulomb Hamiltonian, and although $h_{x_{1}}$ contains a short-range potential, it does not support bound states. Figure I shows the analytic structure on the integrand (\ref{contourint})
and the contour which encircles the spectrum of $h_{y_{1}}$ without penetrating into the
spectrum of $h_{x_{1}}$. In Fig.\ II the contour $C$ is deformed analytically. This is a
more advantageous contour for numerical calculations since the matrix elements of Green's operators falls off quickly and smoothly in this complex imaginary direction.

\begin{figure}[th]
\centering
\includegraphics[width=10cm]{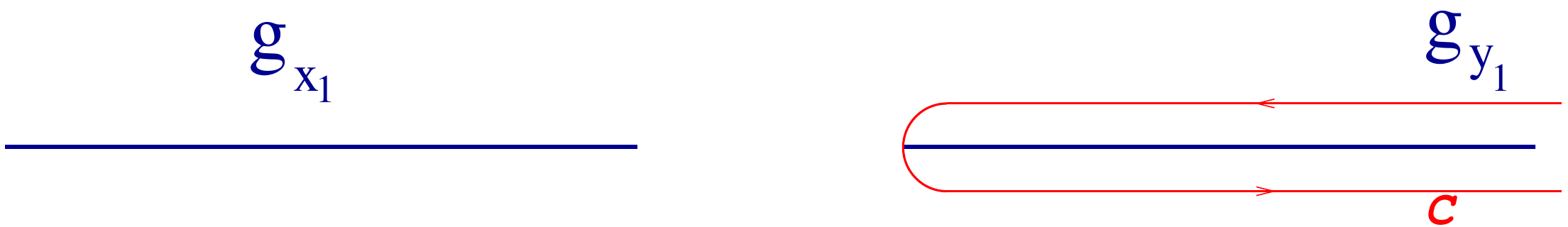}
\vspace*{8pt}
\caption{A schematic illustration of the analytic structure of the integrand (\ref{contourint}).
If $E<0$, the two branch-cuts are well separated. The contour $C$ goes around the spectrum of
$h_{y_{1}}$ without penetrating into the spectrum of $h_{x_{1}}$.\label{f1}}
\end{figure}

\begin{figure}[th]
\centering
\includegraphics[width=10cm]{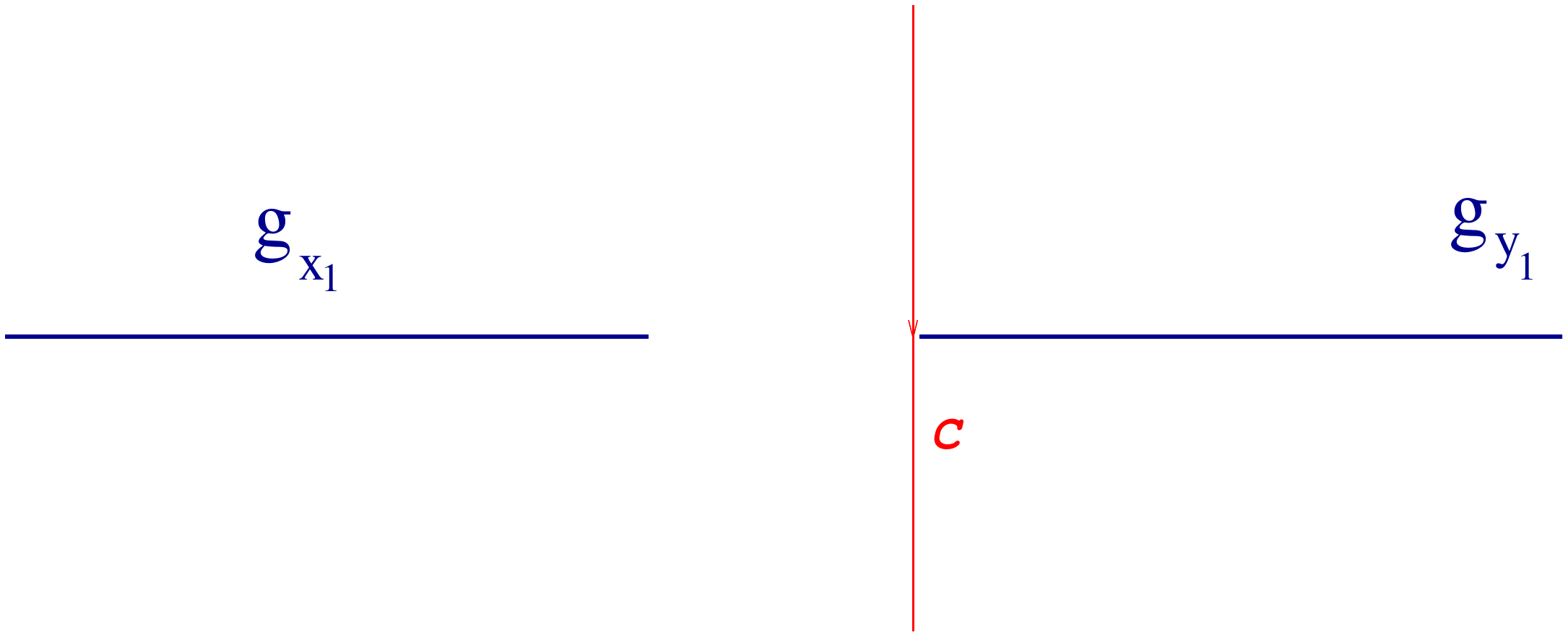}
\vspace*{8pt}
\caption{The same as in  Fig.\ 1. Here the contour is deformed analytically to achieve
a smother integrand.\label{f2}}
\end{figure}

\section{The $\alpha-\alpha$ potential}

Two- and three-$\alpha$ systems have been subject to a very intensive study over the past decades.
We just refer here to some recent investigations ranging from local plus three-body potential models\cite{filikhin}, through orthogonality condition model\cite{3ocm}, to genuine resonating group calculations\cite{resgroup}.

\subsection{The Ali-Bodmer potential}

The ''standard model'' for the $\alpha-\alpha$ interaction is the Ali-Bodmer potential\cite{alibodmer}. There are various parameterizations but, in general, in this model, the potential is local and partial-wave dependent. We use the parametrization
\begin{equation}
V_{AB}(r)=
\begin{cases}
-150 \exp(-0.5 r^{2}) + 1050 \exp(-0.8 r^{2})& \text{if $l=0$} \\
-150 \exp(-0.5 r^{2}) + 640 \exp(-0.8 r^{2}) & \text{if $l=2$} \\
-150 \exp(-0.5 r^{2})  & \text{if $l \ge 4$ even},
\end{cases}
\end{equation}
which is shown in Fig.\ III. We can see that the potential has a very strong angular momentum
dependence.

\begin{figure}[th]
\centering
\includegraphics[width=12cm]{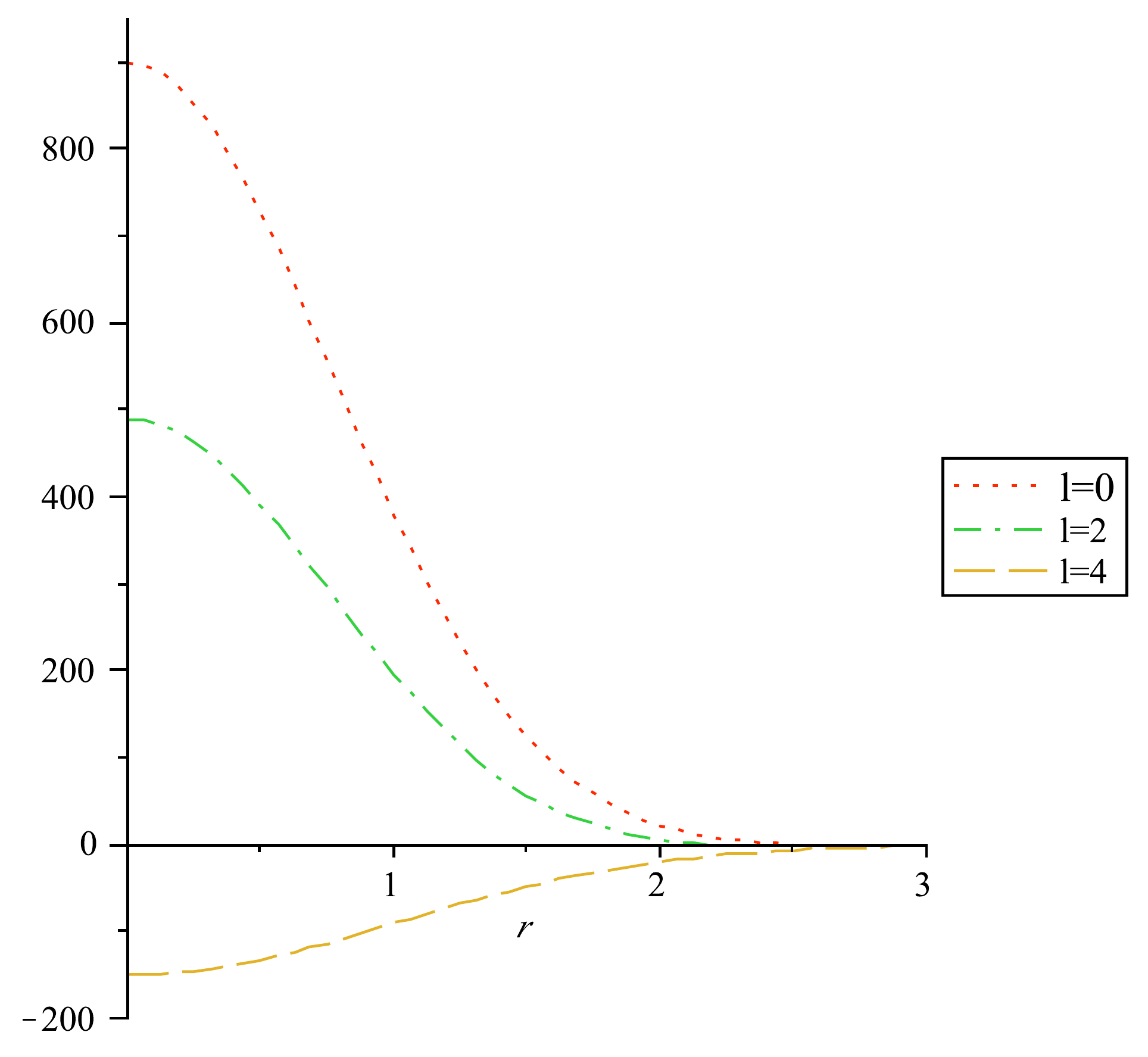}
\vspace*{8pt}
\caption{The Ali-Bormer $\alpha-\alpha$ potential for $l=0$, $l=2$ and $l=4$.\label{f3}}
\end{figure}

 Table I shows the binding energy of the three-$\alpha$ system with total angular momentum $L=0$ as 
a function of maximal subsystems angular momenta $l_{max}=\lambda_{max}$. We used CS function up to $N=25$, which provided us with about five digits accuracy.
 If we take the $l=\lambda=0$ angular momentum channel only, we do not get binding.
 We can also see that angular momentum channels $l=\lambda > 4$ have very little
 contributions to the binding energy. The experimental binding energy of the three-$\alpha$ ($C_{12}$) system is $E_{3\alpha}=-7.275$ MeV, so the Ali-Bodmer potential does not provide
 enough binding.

\begin{table}[pt]
\tbl{$L=0$ three-$\alpha$ binding energy as a function of subsystem angular momentum $l_{max}$ in case
if Ali-Bodmer (AB) potential, fish-bone potential of Kircher and Schmid (FB-1) and the results of this work
(FB-2).}
{\begin{tabular}{@{}ccccc@{}} \toprule
$l_{max}$ & AB & FB-1 & FB-2 \\ \colrule
2 &  -1.70 &  0.057 & -0.313 \\
4 &  -2.26 & -15.47 & -7.112 \\
6 & -2.28  & -15.63 & -7.273 \\
8 & -2.28  & -15.63 & -7.275 \\ 
\botrule
\end{tabular} }
\end{table}

\subsection{The fish-bone optical model}

The fish-bone model is motivated by the cluster model. In the resonating group model the total wave function is an antisymmetrized product of the cluster $\Phi$ and the inter-cluster $\chi$ relative states
\begin{equation}
|\Psi \rangle = |\{{\cal A} \Phi \chi \}\rangle.
\end{equation}
The state $\Phi$, which is supposed to be known in this model, describes the internal properties of the clusters, including spin and isospin structure. 
The unknown relative motion state $\chi$ is determined from the variational anzatz
\begin{equation}
\langle \Phi \delta \chi | {\cal A} (H-E) {\cal A} | \Phi \chi \rangle =0.
\end{equation}
This anzatz results in a rather complicated equation for $\chi$, which were possible to solve only by using serious approximations on $\Phi$ and on the interaction of the particles. Based on this resonating group cluster model, several models has been proposed to describe the motion and interaction of composite particles. Probably the most elaborated one is the fish-bone optical model proposed by Schmid\cite{schmid}.

In the fish-bone model the interaction of the two-body subsystem is given by 
\begin{equation}
{\cal V}_{l}=v_{l}-\sum_{i,j}| u_{l,i}\rangle \langle u_{l,i}|(h^{0}+v_{l}-\epsilon_{l,i})|u_{l,j}\rangle \bar{M}_{l,ij}\langle u_{l,j}|,
\end{equation}
where $l$ refers to partial wave, $h^{0}$ is the kinetic energy and $v_{l}$ is a local potential. The states $|u_{l,i}\rangle$ are eigenstates of the norm operator,
\begin{equation}
\langle \Phi \vec{r} | {\cal A} | \Phi u_{l,i} \rangle = (1-\eta_{l,i})\langle \vec{r} |u_{l,i}\rangle,
\end{equation}
where $\vec{r}$ is the center of mass distance of the two clusters. If the relative motion is forbidden by Pauli principle then $\langle \Phi \vec{r} | {\cal A} | \Phi u_{l,i} \rangle=0$,
and $\eta_{l,i}=1$. The $\eta_{l,i}$ eigenvalues are ordered such that $|\eta_{l,i}| \ge |\eta_{l,i+1}|$. The matrix $\bar{M}$ is then given by
\begin{equation}
\bar{M}_{ij}=
\begin{cases}
1-\frac{ \displaystyle 1-\eta_{l,i}}{ \displaystyle [(1-\bar{\eta}_{l,i}) (1-\bar{\eta}_{l,i})]^{1/2}}, & \text{if $i \le j$}, \\
 1-\frac{ \displaystyle  1-\eta_{l,j}}{ \displaystyle   [(1-\bar{\eta}_{l,j}) (1-\bar{\eta}_{l,i})]^{1/2}}, & 
 \text{if $i > j$},
 \end{cases}
\end{equation}
where $\bar{\eta}_{l,i}=0$ if $\eta_{l,i}=1$ and $\bar{\eta}_{l,i}=\eta_{l,i}$ otherwise.  Or, in matrix form, if we have one Pauli forbidden state,
\begin{equation}
\bar{M}_{l}=\left(\begin{matrix}
1 & 1 & 1 & 1 & \ldots \\
1 & 0 & 1-\sqrt{\frac{1-\eta_{l,2}}{1-\eta_{l,3}}} & 1-\sqrt{\frac{1-\eta_{l,2}}{1-\eta_{l,4}}} &  \ldots \\
1 & 1-\sqrt{\frac{1-\eta_{l,2}}{1-\eta_{l,3}}} & 0 & 1-\sqrt{\frac{1-\eta_{l,2}}{1-\eta_{l,4}}} &   \ldots \\ 
1 & 1-\sqrt{\frac{1-\eta_{l,2}}{1-\eta_{l,3}}} & 1-\sqrt{\frac{1-\eta_{l,2}}{1-\eta_{l,4}}} & 0 &   \ldots \\ 
\vdots & \vdots & \vdots &\vdots & \ddots  \\
\end{matrix}\right),
\end{equation}
which exhibits a fish-bone-like structure, where the name of the model comes from. In this model the Pauli-forbidden states become eigenstates at $\epsilon$ energy. By choosing $\epsilon$ as large positive, they become bound states at large positive energy, and thus disappear from the physically relevant part of the spectrum. There are several versions of the fish-bone model which differ in off-shell transformations, i.e.\ in transformations which effect the internal part of the wave function and leave the asymptotic part, and the spectrum, unchanged. This version of the model minimizes the three-body potential, which is therefore neglected.

We assume that in the $\alpha$ particles the nucleons are in $0s$ states in an oscillator well of width parameter $a$. Then the norm kernel eigenvalues are also harmonic oscillator functions with the same width parameter and the eigenvalues are known\cite{horiouchi}: $\eta_{0,i}=1,1,1/4,1/16,1/64,\ldots$, $\eta_{2,i}=1,1/4,1/16,1/64,\ldots$ and $\eta_{4,i}=1/4,1/16,1/64,\ldots$.  So, in the $l=0$ relative motion channel there are two Pauli-forbidden states, in $l=2$ there is one, an in $l=4$ and higher channels there are none. The decreasing value of $\eta$ indicates that those harmonic oscillator sates in the relative motion are less and less suppressed by the Pauli principle. For the $\epsilon$ parameter of the fish-bone model, which aim is to remove the Pauli-forbidden states, we took $\epsilon=200000$ MeV. In this range of $\epsilon$, the dependence of the results was beyond the fifth significant digit.
 
In an earlier work by Kircher and Schmid\cite{kircher} a fish-bone-type potential was determined from two-$\alpha$ scattering data. The width parameter was fixed to $a=0.55 \mbox{fm}^{-2}$ and the  the local potential was given by
\begin{equation}
v_{l}(r)=v_{0} \exp(-\beta r^{2})+\frac{4\mbox{e}^{2}}{r} \mbox{erf}\left( \sqrt{\frac{2a}{3}} r \right)~,
\end{equation}
where $v_{0}=-108.41998 \mbox{MeV}$ and $\beta= 0.18898 \mbox{fm}^{-2}$. While 
this potential provides a reasonably good fit to $l=0$, $l=2$ and $l=4$ partial wave phase shifts, it seriously
overbinds the three-$\alpha$ system (see FB-1 in Table I). One may conclude that there is a need for three-body potential. This was the choice Oryu and Kamada\cite{oryu} adopted. They added a phenomenological three-body potential to the fish-bone potential of Kircher and Schmid and found that a huge three-body potential is needed to reproduce the experimental data.
But, our Faddeev calculations reveal that the $l=4$ partial wave is very important to the three-$\alpha$ binding and, for this partial wave, the fit to experimental data is not so stellar. So we concluded, that it may be possible to improve the agreement in the $l=4$ partial wave and achieve a better description for the three-$\alpha$ binding energy.

As a local potential we took two Gaussians plus screened Coulomb potential
\begin{equation}
v_{l}(r)=v_{1} \exp(-\beta_{1} r^{2})+v_{2} \exp(-\beta_{2} r^{2})+\frac{4\mbox{e}^{2}}{r} \mbox{erf}\left( \sqrt{\frac{2a}{3}} r \right),
\end{equation}
where $v_{1}$, $\beta_{1}$, $v_{2}$, $\beta_{2}$ and $a$ are fitting parameters. In the fitting procedure we incorporated the famous $^{8}Be$, $l=0$ resonance state at $E_{2b}^{exp}=(0.0916-0.000003{\rm i})$ MeV, the $^{12}C$ three-$\alpha$ ground state energy $E_{3b}^{exp}=-7.275$ MeV, and the $l=0$, $l=2$ and $l=4$ low energy phase shifts. With parameters
$v_{1}=-120.30683493$ MeV,  $\beta_{1}=0.20206127\ {\rm fm}^{-2}$,  $v_{2}=49.06187648$ MeV,    $\beta_{2}=0.76601097\ {\rm fm}^{-2}$ and   $a=0.64874009\ {\rm fm}^{-2}$ we achieved a perfect fit. For the $l=0$ two-body resonance state we get $E_{2b}=0.09161  -0.00000303 {\rm i}$ MeV, and for the three-body ground state $E_{3b}=-7.27502$ MeV. The fit to the phase shifts is shown on Fig 4. We can see that the fit could hardly be any better.
Fig 5 shows the local part of the fish-bone potential. Notice that unlike with the Ali-Bodmer potential, we achieved this agreement by using the same potential in all partial waves. 

\begin{figure}[th]
\centering
\includegraphics[width=12cm]{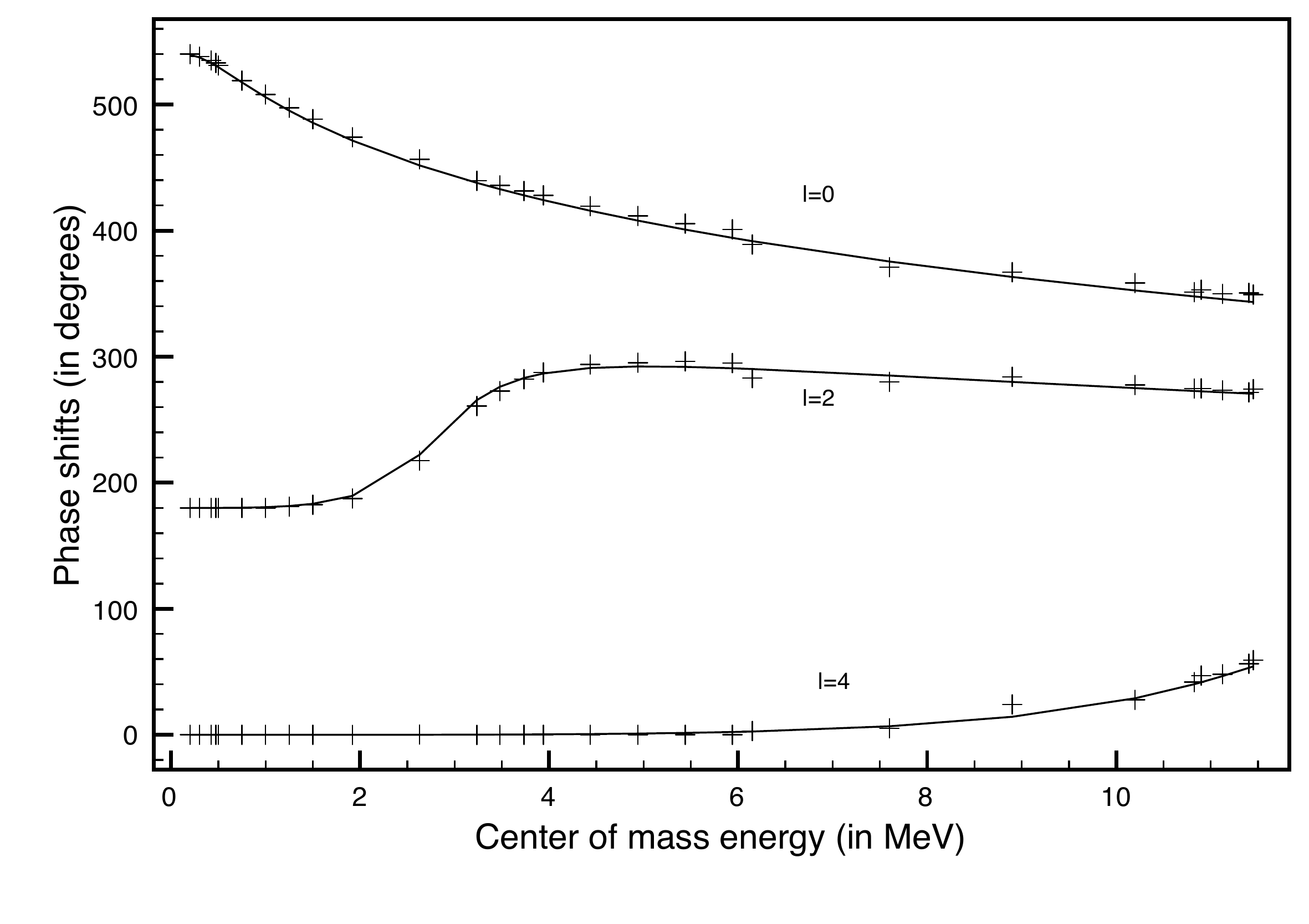}
\vspace*{8pt}
\caption{Fit to the experimental $l=0$, $l=2$ and $l=4$ phase shifts.\label{f4}}
\end{figure}

\begin{figure}[th]
\centering
\includegraphics[width=12cm]{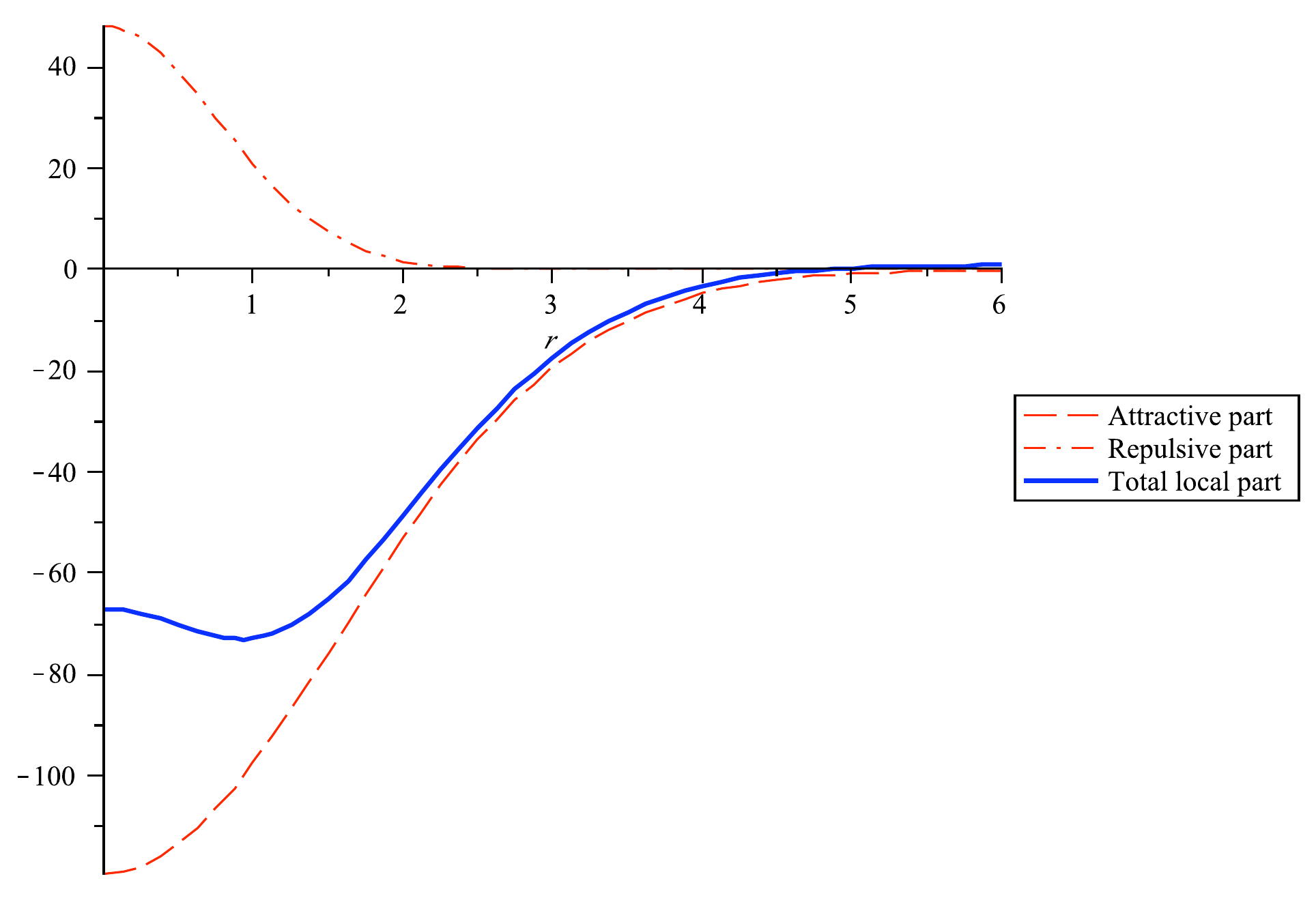}
\vspace*{8pt}
\caption{The local part of the fish-bone potential $v_{l}$.\label{f5}}
\end{figure}

Having this new $\alpha-\alpha$ fish-bone potential from the fitting procedure, we also calculated the first excited state of the three-$\alpha$ system. This state is a resonant state, and we got $E_{3\alpha}^{res}=(0.54 -0.0005{\rm i})$ MeV, which is again very close to the experimental value.

\section{Summary}

In order that we can incorporate the Pauli effect we have to leave the local potential models and have to use nonlocal potentials.  We solved the underlying Lippmann-Schwinger and Faddeev integral equations by performing a separable expansion of the potential in Coulomb-Sturmian basis. This expansion scheme can handle practically any kind of Coulomb plus nonlocal short-range potentials. The Faddeev equations have a further advantage that they allow a very detailed channel-by-channel analysis of the interaction. The method is also very efficient numerically. To solve the Faddeev equation for five angular momentum channels takes a few minutes on our Mac PC, and to fit the fish-bone potential parameters took a couple of hours. We believe that we possess the right model and appropriate tools for investigating the interaction of composite particles.

It is truly remarkable that we got a unified description of the two-$\alpha$ $l=0$ resonant state and the $l=0$, $l=2$ and $l=4$ phase shifts, as well as the three-$\alpha$ binding energy. This was achieved without resorting to three-body potentials and adding explicit angular momentum dependence to the local potential. The results show that the angular momentum dependence comes form the Pauli effect. 
If the Pauli effect is properly incorporated, like in the fish-bone model, and the fit is done in such a way that both two- and three-body data are incorporated, there is no need for additional angular momentum dependence and there is no room for a strong three-body potential. We can see that the proper inclusion of the Pauli principle simplifies the model potential of composite particles. 

\section*{Acknowledgments}

The authors are indebted to E.\ W.\ Schmid and R.\ G.\ Lovas for useful discussions. This work has been supported by the Research Corporation.

\end{document}